\documentclass[aps,prd,12pt]{revtex4}
\usepackage{amssymb}
\usepackage{amsmath}
\def\journal#1#2#3#4{{#1} {\bf #2}, #3 (#4)}
\newcommand{\be}{\begin{equation}}
\newcommand{\ee}{\end{equation}}
\newcommand{\bea}{\begin{eqnarray}}
\newcommand{\eea}{\end{eqnarray}}
\newcommand{\hf}{\frac12}
\newcommand{\nn}{\nonumber\\}
\def\eq#1{(\ref{#1})}

\def\hphi{{\hat\phi}}
\def\hD{{\hat D}}
\def\mr#1{{\mathrm{#1}}}
\def\ord#1{{\cal O}\left(#1\right)}
\def\fdd#1#2#3{\frac{\delta^2#1}{\delta#2\delta#3}}
\def\pd#1#2{\frac{\partial#1}{\partial#2}}

\begin{document}
\title{Stability and causality of multi-local theories}
\author{Janos Polonyi}
\affiliation{Strasbourg University, High Energy Theory Group, CNRS-IPHC,23 rue du Loess, BP28 67037 Strasbourg Cedex 2 France}
\date{\today}

\begin{abstract}
The regularized theories are non-local at the scale of the cutoff, leading so to the usual difficulties of non-local theories. In this work the conservation laws and causality are investigated for classical field theories with multi-cluster action. The conservation laws are found to play a less significant role than in local theories because due to the non-locality the conserved quantities are not integrals of the motion, and they can exist even without underlying symmetries. Moreover, the conservation of the energy can not prevent instability brought about by the unbounded nature of the energy from below. Hence a sufficient condition of stability is lost. Theories, obtained by appropriate point splitting of local interactions are shown to be causal thereby a necessary condition of stability can be retained.
\end{abstract}
\maketitle

\section{Introduction}
While the fundamental laws of physics are supposed to be local, simplifying thereby their treatment enormously the observations, carried out always with limited precision, can not follow all degrees of freedom. Hence we explore the effective dynamics of the resolved components and that is non-local. This non-locality, generated by eliminating degrees of freedom from a local dynamics, would be a pure technical issue if one could return to the full, local description. There is however a genuine physical problem here, owing to the lack of finite local field theories. Our local theories are plagued by UV divergences and can only be made well defined by the help of a cutoff, characterizing our ignorance, our inability to resolve all scales in the observations. The regulated cutoff theories are genuinely non-local at the scale of the UV cutoff without the possibility of returning to a local dynamics. This problem motivates a more careful study of non-local theories, irrespectively whether they are fundamental or effective. 

Non-locality in time opens up the possibility of loosing two important features of our theories. First of all, it leads to instability as a rule rather than an exception. This can be understood in the simplest way by noting that non-local interactions in time violate Newton's third law. In fact, let us assume that interaction energy between two particles, described by the coordinates $x$ and $y$, is given by $U(x(t)-y(t'))$, containing the coordinates at different times, $t\ne t'$. The forces acting on the coordinates $x$ and $y$, $F_x(t)=-\nabla U(x(t)-y(t'))$ and $F_y(t')=\nabla U(x(t)-y(t'))$ do not cancel out in calculating the change of the total momentum at a given time, and the overwhelmingly large phase space at high energies makes the motion unstable. Another feature, endangered by non-locality in an obvious manner, is the causality. The question we embark here is whether the non-local feature of a dynamics is compatible with stability and causality. No general theorem is found to assure the stability but a family of causal non-local theories is proposed which may support stable and causal dynamics.

The winding road of non-local theories started with an attempt to establish an UV finite perturbation expansion \cite{uvfinite}, see ref. \cite{krizhnits} for a review of the early results. The non-local interactions open up a series of severe problems: the violation of the causality \cite{causality}, the spoiling of gauge invariance \cite{gauge}, and the emergence of instability \cite{instab,eliezer}. The non-locality makes necessary to generalize the Hamiltonian \cite {hamiltonian} and the variational  \cite{marnelius,mita,huang,kegeles,variational} formalisms, too. A special attention will be paid below to non-localities, arising from the point splitting \cite{psplitting} since that seems to be the only non-perturbative, relativistic regulator which may produce stable dynamics \cite{cer}.

The non-local action functionals are rather involved mathematical objects hence we narrow down the discussion into a physically justifiable, simpler functional space. We follow the guidance of perturbative effective theories and restrict our attention to to multi-local actions, the sum of multi-cluster terms, the $n$-cluster contribution being an $n$-fold space-time integral of an $n$-cluster Lagrangian, c.f. eq. \eq{multilac} below.

The goal of the present paper is to point out three properties of classical multi-local theories: a modified status of conservation laws, namely their independence of symmetries, the reduced usefulness of the energy conservation and finally a possibility of constructing causal models by the help of point splitting.  The stability is discussed by regarding the classical energy expression, the issue of causality is presented in a manner which is valid both for classical and for quantum systems.

The problem of the uniqueness of a multi-local action and the corresponding conservation laws has already been considered in ref. \cite{marnelius}. Noether's theorem has been derived in bi-local field theories \cite{mita}, in non-local elasticity \cite {huang} and in group field theories \cite{kegeles}, as well. The present work addresses  different features of the conservation laws, namely their relation with balance equations and the symmetries furthermore the usefulness of the energy conservation.

The energy conservation plays a special role among the conservation laws, inasmuch as used to control the stability. However one should keep in mind that this holds only in theories with bounded energy from below. A characteristic feature of the multi-local dynamics is that the definiteness of the conserved current densities is lost, namely the energy, derived from a time translation invariant multi-local action, is non-definite and unbounded, allowing a conservative dynamics be unstable.

The boundedness of the energy from below is a sufficient but not necessary condition of the stability, there might exists other bounded conserved quantities which keep the dynamics stable \cite{kaparulin}. Hence it is important to check explicitly if there are unbounded, runaway solutions of the equation of motion. A necessary condition of the boundedness of the trajectories is causality. The possibility of trading instability into acausality has already been noticed in relation to the Abraham-Lorentz force, \cite{dirac}. This is actually a rather general phenomenon, well beyond electrodynamics, owing to the tacit assumption about the boundedness of the trajectory in using the Fourier transformation. In fact, the time dependence of the retarded Green function, obtained by the formal application of the residue theorem,
\be\label{instacaus}
G^r(t)=2\pi i\sum_{\omega_p}\Theta(-t~\mr{Im}\omega_p)Res[\tilde G^r(\omega_p)]e^{-i\omega_pt},
\ee
is always bounded, containing the runaway modes with $\mr{Im}\omega_p>0$ for $t<0$ only.

The non-local dynamics not only makes the causality a non-trivial issue, it necessitates the imposition of further initial conditions \cite{barnaby}. Such a proliferation of the auxiliary conditions can be understood as a transmutation of the initial conditions of the unobserved environment into the effective dynamics. This difficulty is circumvented below by starting with trivial, static motion at a sufficiently early initial time, $t_i\to-\infty$, and driving the system to the desired initial state by the help of an external source.

While the pole structure of the retarded Green's function is left to be rearranged by hand within the usual Lagrangian formalism there is an extension of the variation principle, the Closed Time Path (CTP) formalism \cite{schw}, generalized for classical systems \cite{effth}, providing a constructive definition of the retarded Green's function. It is shown below within the framework of this formalism that multi-local theories, originating from point splitting , can be extended within the CTP scheme in a causal manner, thereby opening the possibility that such non-local theories remain stable at least for certain choice of their parameters.

\section{Multi-local actions}
While the fundamental physical laws, governing closed systems, are assumed to be local, the effective equations of motion of an open system are non-local. One can easily follow the build up of the non-local feature perturbatively in a toy model where the coordinates $x$ and $y$ belong to the observed system and its environment, respectively and the closed, conservative dynamics of the full system is governed by a local action, $S[x,y]=S_s[x]+S_e[x,y]$. The environment action is written in the form
\be
S_e[x,y]=\int dt\left[\hf yD^{-1}y-yk(x)-U(x,y)\right],
\ee
$k(x)$ and $U=\ord{y^2}$ representing the system-environment interactions. The environment initial conditions are chosen to be trivial, $y$ and its time derivatives being vanishing at $t_i=-\infty$. In theories without linear, $\ord{xy}$ interactions the unstable solution, $y=0$, is avoided by introducing an $x$-independent external source, $k(x)=k$, with suitable chosen time dependence to drive the environment adiabatically to a non-trivial state. 

The effective action for the system is obtained by inserting the solution of the environment equation of motion for a general system trajectory, $y=y[x]$, into the action, $S_{eff}[x]=S[x,y[x]]$. The iterative solution of the environment equation of motion, $y^{(n+1)}=D^r[\partial_yU(x,y^{(n)})+k(x)]$, the poles of the retarded Green's function, $D^r$, being shifted to the physical sheet in the complex frequency plane, can be represented by a series of tree graphs. The $n$-th order graph contains $n$ vertices, $\partial_yU(x,y)$, connected by $D^r$, and the factors $x$, appearing at the vertices and at the the source $j(x)$,  represent the external legs. 

The effective action, obtained by this solution is multi-local, it can be written as the sum of $n$-cluster contributions, with independent integration over the time of the local clusters. Note that the translation invariance in time allows the dependence of the integrand of a multi-cluster term on the time differences of the clusters. The latter, a memory effect, is generated by the environment Green's function and reflects an important scale dependence of the effective dynamics. Such a procedure yields a multi-local action for a generic real scalar field, $\phi(x)$, 
\be\label{multilac}
S=\sum_{1\le n}^\infty\frac1{n!}\int_Vdx_1\cdots dx_nL^{(n)},
\ee
with a translation invariant Lagrangian, 
\be\label{multilacl}
L^{(n)}=L^{(n)}(x_1,\phi(x_1),\partial\phi(x_1),\ldots,x_n,\phi(x_n),\partial\phi(x_n)),
\ee
which is symmetric under the permutation of the cluster locations for $n\ge2$ and depends on $x_m-x_{m'}$. To avoid Ostrogradsky's instability \cite{ostrogadsky}, we restrict our attention to action functionals, depending at most on the first time derivatives of the fields and furthermore the dependence of the contributions to the action on the cluster separation in time is assumed to be sufficiently long range to apply the usual variational calculus in terms of the fields and their first derivatives.

The two-cluster form of the effective dynamics of the electric charges has been known since long time \cite{sctefowf}. This procedure generates conservative forces only, the construction of nonconservative effective dynamics, in particular the radiation field, requires the use of the classical CTP formalism \cite{effth}. 

Note that the tree graphs, generated by the solution of the classical environmental equation of motion, are turned into graphs with loops by the elimination of the environment. In fact, it is well known ever since the Rayleigh-Jeans law that the dynamics of the classical electromagnetic field, interacting with a thermal environment, is spoiled by UV divergences. This is due to such loop diagrams of the classical effective action which generate UV divergences and make the introduction of a regulator and the use of the renormalization group method necessary \cite{cer}.

A similar, multi-local action arises in the quantum case, as well, where the the perturbative elimination of the environment coordinate leads to the environment propagator, as internal link, connecting the clusters of the Feynman graphs, contributing to the effective action. 

The dynamics of certain multi-local theories can be localized by the introduction of additional space-time dimensions \cite{calcagni}. While such a modification of the theory renders the UV divergences more serious it would be interesting to see whether this idea can be used to render perturbative effective theories more  local.

\section{Conservation laws}
A system of $N$ one dimensional classical particles, obeying a local, autonomous dynamics displays $2N-1$ independent constants of motion, i.e., functions, $C_j(x(t),\dot x(t),t)$, which are local in time and remain constant along the physical trajectories. It is customary to define two overlapping subsets of the constants of motion, the conserved quantities and the integrals of motion. The former set is related to continuous symmetries and plays a more important role in the dynamics \cite{landaum} and the latter consists of the constants of motion without explicit time dependence. The conserved quantities are defined by Noether's theorem and can be recast in the form of an integral of motion by the help of the equation of motion. Thus the set of integrals of motion contain the conserved quantities. However the opposite is not true, a set of continuous, time-independent transformations which do not leave the equation of motion invariant can still be used to derive a balance equation, 
\be\label{balance}
\partial_tQ=\rho,
\ee
which becomes an integral of motion after the elimination of $\ddot x$ by the help of the equation of motion. The simplicity of the integrals of motion makes them important by allowing to derive further integrals of motion in an algebraic manner, without solving the equation of motion. There might be some further important constants of motion with simple time dependence, for instance the Runge-Lenz vector in Coulomb problems. 

The situation is fundamentally different in multi-local dynamics. (i) An important property of the local dynamics, the absence of the explicit time-dependence of the equation of motion, does not appear to be a natural concept and the autonomy and the time translation invariance, two equivalent properties of local dynamics, become different. In fact, the perturbative elimination of the degrees of freedom from an autonomous dynamics generates a non-autonomous, time translation invariant effective dynamics, cf. \eq{multilacl}. (ii) One can not identify the trajectories by specifying some initial conditions hence the number of constants of motion, i.e., time-independent, multi-local functionals of the trajectory, is unknown. (iii) The explicit appearance of time in the Lagrangian of a time translation invariant dynamics, cf. \eq{multilacl}, eliminates the integrals of motion from the family of the constants of motion. (iv) Finally, the relation between the conserved quantities and the symmetries is lost. On the one hand, the continuous symmetries do not generate conserved quantities: Noether's theorem, worked out below for multi-local dynamics in field theory, leads to a balance equation \eq{balance} rather than conservation law. On the other hand, the conservation laws need no symmetry. In fact, let us suppose that we have a set of transformations, forming a continuous manifold and containing the identity. One can easily derive a balance equation \eq{balance} by following the strategy of Noether's theorem. The modification of $Q$,
\be\label{balcons}
Q(t)\to Q'(t)=Q(t)-\int_{t_i}^tdt'\rho(t'),
\ee
transforms a balance equation into a conservation law. One can show that the violation of a continuous symmetry can be compensated by an appropriately defined multi-local term. Such a trivial modification is possible both in local and non-local dynamics however is not allowed in the former case where $Q$ is supposed to be local. The change \eq{balcons} is allowed in a multi-local dynamics, the only special feature being that the compensating term contains one more cluster than the action. Points (ii)-(iv) may loosely be summarized by saying that there are more conservation laws in multi-local dynamics, unrelated to continuous symmetries, however their usefulness is somehow limited, being a multi-local functional of the trajectory.

We briefly review the new features of the conservation laws, mentioned above, for a multi-local field theory. The action for a real, multi component scalar field, $\phi(x)$, is written in the form \eq{multilac}-\eq{multilacl} where the dependence on $x_m-x_{m'}$ is assumed to be sufficiently long range to avoid Ostrogradsky's instability \cite{ostrogadsky}. The equation of motion,
\be\label{sceom}
0=\sum_{n\ge1}\frac1{(n-1)!}\int_Vdx_2\cdots dx_n\left(\pd{L^{(n)}}{\phi_1}-\partial_\mu\pd{L^{(n)}}{\partial_\mu\phi_1}\right),
\ee
where $\phi_j=\phi(x_j)$ and the derivatives with respect to the field variables $\phi_j$, $j\ge2$ reduce to the same expression owing to the symmetry of the Lagrangian $L^{(n)}$ with respect to the clusters exchange. The point splitting \cite{psplitting} consists of smearing the interactions in the space-time. If the interaction is described by a local potential energy, $U(\phi(x))$, then its smeared version, $U(\tilde\phi(x))$, is defined by the help of the smeared field variable,
\be\label{smearing}
\tilde\phi(x)=\int_V dy\chi(x-y)\phi(y).
\ee
The smearing function, $\chi(x)$, depends on the cutoff, $\Lambda$, in such a manner that $\chi(x)\to\delta(x)$ as $\Lambda\to\infty$. The non-locality of the regulated theory, appearing at the scale of the cutoff, is due to the UV environment and is represented by a phenomenological form of the smearing function, $\chi(x)$. The point splitted action of a scalar field theory with a local, $\ord{\phi^N}$ self-interaction energy $U(\phi)$,
\be\label{pspla}
S=\int_Vdx\left[\hf\partial\phi(x)\partial^\mu\phi(x)-\frac{m^2}2\phi^2(x)-U(\tilde\phi(x))\right],
\ee
contains $N$-cluster terms.

Let us consider a linear internal space transformation group, $\phi(x)\to g\phi(x)$, whose infinitesimal transformations, $\phi(x)\to\phi(x)+\delta\phi(x)$, $\delta\phi(x)=\epsilon\tau\phi(x)$, $\tau$ being a real matrix, which induces the change, 
\be\label{changeint}
r^{(n)}=\sum_{j=1}^n\left(\pd{L^{(n)}}{\phi_j}\tau\phi_j+\pd{L^{(n)}}{\partial_\mu\phi_j}\tau\partial_\mu\phi_j\right),
\ee
of the Lagrangian, and assume that the field configuration, $\phi(x)$, satisfies the equation of motion. We treat $\epsilon$ as a space-time dependent variational parameter, controlled by the linearized action,
\be
S[\epsilon]=\sum_n\frac1{(n-1)!}\int_Vdx_1\cdots dx_n\left(\epsilon_1\frac{r^{(n)}}n+\partial_\mu\epsilon_1\pd{L^{(n)}}{\partial_\mu\phi_1}\tau\phi_1\right),
\ee
where the equation of motion was used to eliminate the second order derivatives of $\phi$. The equation of motion for $\epsilon$ can be written in the form of a balance equation, $\partial_\mu j^\mu=\rho$, with the Noether current,
\be\label{currio}
j^\mu(x_1)=\sum_n\frac1{(n-1)!}\int_Vdx_2\cdots dx_n\pd{L^{(n)}}{\partial_\mu\phi_1}\tau\phi_1,
\ee
and the source,
\be\label{sourceio}
\rho(x_1)=\sum_n\frac1{n!}\int_Vdx_2\cdots dx_nr^{(n)}.
\ee

It is known from the solution of Gauss' law in classical electrodynamics that any scalar can be written as a divergence by allowing non-local expressions. In a similar manner the modified current,
\be
j'^\mu(x)=j^\mu(x)+\int dy\partial^\mu D_0^r(x-y)\rho(y).
\ee
where $D^r_0$ stands for the massless retarded Green's function satisfying the equation $\Box D^r=-1$, is conserved. The emergence of a conserved current without a symmetry can be understood by a non-local extension of the theory where a new field, $\psi(x)$, is introduced to implement an invariance with respect to the transformations. The action of the extended model,
\be
S=\sum_n\frac1{n!}\int dx_1\cdots dx_ndx_{n+1}L^{(n)}(x_1,g(\psi_{n+1})\phi_1,\partial g(\psi_{n+1})\phi_1,\ldots),
\ee
is one order higher in the cluster expansion than the original, respects the global symmetry, $\phi(x)\to\phi'(x)=g(z)\phi(x)$, $\psi(x)\to\psi'(x)$ with $g(\psi'(x))=g(\psi(x))g^{-1}(z)$, and yields the same equations of motion for $\phi(x)$ as the original model. The transformation of $\psi$ requires that the set of transformation form a group. It is an easy exercise to show that the balance equation with the current \eq{currio} and the source \eq{sourceio} follows when Noether's theorem is applied to the extended model.

The change of the Lagrange density of the action \eq{pspla} under the internal space transformation,
\be
r(x)=\partial_\mu\phi(x)\tau\partial^\mu\phi(x)-m^2\phi(x)\tau\phi(x)-\partial_\phi U(\tilde\phi(x))\int dy\chi(x-y)\tau\phi(y).
\ee
leads to a balance equation for the current,
\be
j^\mu(x)=\partial^\mu\phi(x)\tau\phi(x)
\ee
and the source term,
\be
\rho(x)=r(x)+\partial_\phi U(\tilde\phi(x))\overleftrightarrow\chi\tau\phi(x),
\ee
where the notation $f(x)\overleftrightarrow\chi g(x)=\int dy[f(x)\chi(x-y)g(y)-f(y)\chi(y-x)g(x)]$ is used. The modified current,
\be\label{modcurr}
j'^\mu(x)=j^\mu(x)+\int dyD_0(x-y)\partial^\mu[r(y)+\partial_\phi U(\tilde\phi(y))\overleftrightarrow\chi\tau\phi(y)],
\ee
is conserved.

As far as the external symmetries are concerned we restrict our attention to translations and rewrite the action in terms of an infinitesimally shifted space-time coordinate, $x^\mu\to x'^\mu=x^\mu+\epsilon^\mu(x)$, 
\be
S=\sum_n\frac1{n!}\int_{V'}\frac{dx_1}{1+\partial\epsilon_1}\cdots\frac{dx_n}{1+\partial\epsilon_n}L^{(n)}(x_1-\epsilon_1,\phi(x_1-\epsilon_1),\partial\phi(x_1-\epsilon_1)\ldots).
\ee
The vanishing of the $\ord\epsilon$ part of the right hand side of this equation,
\be
0=\sum_n\frac1{(n-1)!}\int_Vdx_2\cdots dx_n\left[\int_Vdx_1(\delta L^{(n)}-\partial\epsilon_1L^{(n)})+\int_{V'-V}dx_1L^{(n)}\right]
\ee
can be written for a field configuration, $\phi(x)$, satisfying the equation of motion in the form
\bea
0&=&\sum_n\frac1{(n-1)!}\int_Vdx_2\cdots dx_n\biggl\{\int_{\partial V}dS_{1\mu}\left(\epsilon_1^\mu L^{(n)}-\pd{L^{(n)}}{\partial_\mu\phi_1}\epsilon_1\partial\phi_1\right)\nn
&&-\int_Vdx_1(\epsilon^\mu_1\partial_{x_1^\mu}L^{(n)}+\partial\epsilon_1L^{(n)})\biggr\}.
\eea
This equation, used with a homogeneous shift, $\epsilon(x)=\epsilon$, for an arbitrary space-time region, $V$, yields the balance equation $\partial_\mu T^{\mu\nu}=\rho^\nu$ with the energy-momentum tensor,
\be\label{emtenso}
T^{\mu\nu}=\sum_n\frac1{(n-1)!}\int_Vdx_2\cdots dx_n\left(\pd{L^{(n)}}{\partial_\mu\phi_1}\partial_\nu\phi_1-L^{(n)}g^\mu_\nu\right),
\ee
where $g^{\mu\nu}$ denotes the flat Minkowski metric tensor and the source is given by
\be\label{sourceemo}
\rho_\mu=-\sum_n\frac1{(n-1)!}\int_Vdx_2\cdots dx_n\partial_{x_1^\mu}L^{(n)}.
\ee
The balance equation can also be recovered by generalizing the action for curvilinear coordinate systems and using the alternative definition $T^{\mu\nu}=2\delta S/\delta g_{\mu\nu}/\sqrt{-g}$. Furthermore, it  can be rewritten in the form of a continuity equation for the modified energy-momentum tensor
\be
T'^{\mu\nu}(x)=T^{\mu\nu}(x)+\int dyD_0^r(x-y)\partial^\mu\rho^\nu(y).
\ee
The energy conservation became a formal equation without any bearing on the stability of the dynamics owing to the unboundedness of the energy from below. 

We have so far considered translation invariant theories. Let us now turn to a non-translation invariant theory and assume for the sake of simplicity that the symmetry breaking arises from the local part of the action. To find a translation invariant extension of the model we replace the local part of the action by the bi-local expression,
\be
S=\int_Vdx_1dx_2L_1(x_1-\psi(x_2),\phi(x_1-\psi(x_2)),
\ee
containing a vector field, $\psi_\mu(x)$, and remaining invariant under the transformation $\psi^\mu(y)\to\psi^\mu(y)+z^\mu$, $x^\mu\to x^\mu+z^\mu$. It is easy to check that the conservation law, corresponding to this symmetry is the local, $n=1$, part of the balance equation with the energy momentum tensor \eq{emtenso} and the source, \eq{sourceemo}.

The action \eq{pspla} yields a balance equation with the energy-momentum tensor,
\be\label{emtensor}
T^{\mu\nu}(x)=\partial^\mu\phi\partial^\nu\phi(x)-\biggl[\hf\partial_\mu\phi\partial^\mu\phi(x)-\frac{m^2}2\phi^2(x)-U(\tilde\phi(x))\biggr]g^{\mu\nu},
\ee
and the source term,
\be
\rho_\mu(x)=\partial_\phi U(\tilde\phi(x))\overleftrightarrow\chi\partial_\mu\phi(x).
\ee
The energy, obtained by integrating $T^{00}(x)$ over the infinite three-space, is non-negative and non-conserved. The modified energy-momentum tensor,
\be\label{memtensor}
T'^{\mu\nu}(x)=T^{\mu\nu}(x)+\int dyD^r_0(x-y)\partial^\mu[U'(\tilde\phi(y))\overleftrightarrow\chi\partial^\nu\phi(y)],
\ee
produces a conserved energy which is however unbounded from below and a sufficient condition of stability is lost.

\section{Causality}
The CTP formalism \cite{schw} offers a systematic and constructive procedure to define the retarded and the advanced Green's functions. To express the classical variational principle for a coordinate $x$ in terms of the initial conditions and to include the non-holonomic, open forces in the classical CTP scheme \cite{effth} one lets the time passes from its initial to its final value, from $t_i$ to $t_f$, followed by a time inversion and the sending the system back to the initial time. The resulting trajectory, $\tilde x(t)$, spanning over the time interval of length $2(t_f-t_i)$ split into a CTP doublet, $\hat x=(x^+,x^-)$, with $x^+(t)=\tilde x(t)$ and $x^-(t)=\tilde x(2t_f-t)$ for $t_i<t<t_f$, describing the forward and the backward moving part of the motion. The double trajectory is closed, namely satisfies the final condition $x^+(t_f)=x^-(t_f)$. The physical trajectory is found by imposing the same initial conditions on $x^\pm(t)$ and solving the variational equations of the action, $S[\hat x]=S[x^+]-S[x^-]$, in the case of a closed system. 

To monitor the dynamics an external source, $j^\pm(t)$, is coupled linearly to $x^\pm(t)$ by the term $\hat j(t)\hat x(t)$ in the action and the two-point Green's functions, making up a $2\times2$ matrix, are defined by the functional derivative,
\be\label{twopgfnct}
D^{\sigma\sigma'}(t,t')=\fdd{S[\hat x]}{j^\sigma(t)}{j^{\sigma'}(t')}_{|\hat j=0},
\ee
evaluated for the trajectory which satisfies the equation of motion and the initial conditions. The action is vanishing for the physical trajectory, $x^+(t)=x^-(t)$, and this degeneracy has to be removed to render the Green's functions well defined. For this end an infinitesimal imaginary part is added to the action, 
\be\label{splitting}
S[\hat x]=S[x^+]-S[x^-]+i\frac\epsilon2\int dt[x^{+2}(t)-x^{-2}(t)].
\ee
It is important to keep in mind that the trajectories of the two time axes, $x^\pm(t)$, are coupled by the final condition, $x^+(t_f)=x^-(t_f)$.

An important advantage of the CTP scheme is that the open systems can be described by an effective CTP action, too. Let us assume that the full system is closed and its dynamics is governed by the action $S[\hat x,\hat y]$, written in terms of the system and the environment coordinates. The effective system action is found by solving the environment equation of motion for a general system trajectory, $\hat y=\hat y[\hat x]$, and inserting it into the original action, $S_{eff}[\hat x]=S[\hat x,\hat y[\hat x]]$. The resulting effective action describes the closed, conservative, as well as the open, non-holonomic forces, the latter being encoded by the couplings between the members of the CTP doublets, $x^+$ and $x^-$. 

It is worthwhile to comment briefly the similarities between the classical and the quantum cases. The particular form of the splitting of the degeneracy in eq. \eq{splitting} allows us to express the two-point function, \eq{twopgfnct}, in terms of three real functions,
\bea\label{grfnctstr}
D^{++}(t)&=&D^n(t)+iD^i(t)=-D^{--*}(-t),\nn
D^{-+}(t)&=&D^f(t)+iD^i(t)=-D^{+-*}(-t),
\eea
and renders the classical and the quantum two-point Green's function of a harmonic oscillator identical, in particular $D^{++}$ becomes the Feynman propagator. The path integral with the integrand $\exp iS[x]/\hbar$ yields the transition matrix element between coordinate eigenstate for closed system and the dependence on the source is used to generate the Green's functions. One can prove in a similar manner that the integral of $\exp iS[\hat x]/\hbar$ over the CTP trajectories, satisfying $x^+(t_f)=x^-(t_f)$, and the initial coordinate being convoluted with the initial density matrix yields the trace of the density matrix where the bra and the ket are evolved in the presence of the external source $-j^-$ and $j^+$, respectively and the functional derivations with respect to $\hat j$ generates the Green's functions.

The real part of the Green's function controls the expectation values, in particular the physical external source, $j^\pm=\pm j$, generates the trajectory
\be
x(t)=\sum_\sigma\int dt'D^{\pm\sigma}(t-t')\sigma j(t'),
\ee
and the block structure \eq{grfnctstr} identifies the retarded Green's function, $D^r=D^n+D^f$. The CTP Green's function is symmetric, implying $D^n(-t)=D^n(t)$, $D^f(-t)=-D^f(t)$, and $D^a(t)=D^r(-t)=D^n(t)-D^f(t)$. The comparison with electrodynamics suggests to identify $D^n$ and $D^f$ with the near and far field Green's functions. The near field describes a dressing of the charge which keeps the energy momentum conserved and is represented by couplings within a time axis. The far field can be identified by the outgoing radiation, representing a non-conservative, dissipative force, acting on the charge and arising from the couplings between the time axis.
  
It is straightforward to generalize the CTP scheme from a point particle to a scalar field theory. The point splitted model is defined by the CTP action,
\be\label{ctppspla}
S[\hphi]=\int dx\left[\hf\hphi(x)\hD^{-1}\hphi(x)-U(\tilde\phi^+(x))+U(\tilde\phi^-(x))\right],
\ee
and the smeared field is given by the convolution,
\be
\tilde{\hphi}(x)=\hat\sigma\int dy\hat\chi(x-y)\hphi(y),
\ee
where $\chi^{ab}(x-y)=\chi^{ba}(y-x)$ and $\hat\sigma=\mr{Diag}(1,-1)$ denotes the ``metric tensor'' of the simplectic structure of the CTP action. The action, written in terms of the smeared field,
\be
S[\tilde{\hphi}]=\int dx\left[\hf\tilde{\hphi}(x)\tilde{\hD}^{-1}\tilde{\hphi}(x)-U(\tilde\phi^+(x))+U(\tilde\phi^-(x))\right]
\ee
contains $\tilde{\hD}=\hat\chi\hD\hat\chi$. We assume that $\tilde\chi$ possess the block structure \eq{grfnctstr} with $\chi^i=0$ to preserve the realness of the field, with $\chi^n=\chi$ to realize the smearing \eq{smearing} and with $\chi^f(x)=\mr{sign}(x^0)\chi^n(x)$ to have causal retarded and advanced components, $\chi^{\stackrel{r}{a}}=\chi^n\pm\chi^f$. The equation $\tilde{\hD}^{\stackrel{r}{a}}=\chi^{\stackrel{r}{a}}D^{\stackrel{r}{a}}\chi^{\stackrel{r}{a}}$ follows from the block structure \eq{grfnctstr} and assures that the dynamics of the smeared field is causal. Note that the theory avoids Ostrogadsky's instability and has a chance to be stable \cite{eliezer} if $\chi^{-1}$, appearing in the kernel of the kinetic energy, $\tilde{\hD}^{-1}=\hat\chi^{-1}\hD^{-1}\hat\chi^{-1}$, is non-polynomial in the momentum space.

\section{Conclusions}
Cutoff theories are intrinsically non-local at the scale of the cutoff and some of the implication of this structure is explored in this work. It is found that the multi-local nature of the dynamics reduces the importance of the conservation laws on two counts. First, the translation invariance in time does not imply  the Lagrangian to be independent of time and hence the generic conserved quantities are not integrals of the motion. Second, Noether's theorem can be used to generate a conserved current even without requiring a symmetry. It is shown for linearly realized internal symmetries and for translations in the space-time that this result follows from a possible extension of the theory, a refinement rather than a coarse graining, defined in such a manner that the new degrees of freedom absorb the violation of the symmetry. 

A distinguished conserved quantity, the energy, plays a unique role among the constants of motion: it serves as an indicator of stability. While the boundedness of the energy from below usually follows trivially from the algebraic structure of a local Lagrangian, such an algebraic argument does not apply anymore to the non-local terms. 

Such a state of affairs renders the usual arguments about the stability of field theories, defined either in the classical or the quantum domain, insufficient. However the point splitting defines an interesting class of models where the original, local field variable is replaced by a smeared, coarse grained field in the interactions in such a manner that the stability might be maintained. The local theories, made non-local by point splitting, can be generalized within the CTP formalism to local, causal models for the coarse grained field. 

The point splitted models which happen to be stable form a realistic set of non-local theories. Numerical results indicate that the effective theory of a point charge, a bi-local theory, belongs to this class \cite{cer}. The clarification of the dynamical origin of the stability remains a primary open problem.

Finally we mention the obvious question whether the instability, found in the classical case, plagues the quantum systems. On the one hand, the quantum theories may be more stable than their classical counterparts, e.g., the stability of a quantum system may be saved by the uncertainty principle if the classical instability is restricted into a small region of the phase space as it happens in the Coulomb problem. On the other hand, if the classical stability is maintained by an energy barrier then the quantum system may be unstable owing to the tunneling. The classical instability of non-local theories arises without any singularity in the action, suggesting that the instability pervades the quantum theory. However the decay time of a false vacuum diverges in the thermodynamical limit, making the stabilization by an energy barrier, if exist,  possible. Dedicated studies are necessary within non-local theories to clarify this point. Such calculations should be carried out in a scheme which leaves the final state unspecified. Within the usual scheme of quantum field theories, based on the transition amplitudes between given states, one imposes fixed initial and final states therefore one needs here the CTP formalism where the initial state is fixed only, to overcome this limitation. Note in this respect that the Feynman-like Green's functions which appear in the usual formalism of quantum field theory are acausal even for a stable system and offer no simple way to diagnose the stability. It remains an intriguing question whether calculations, performed within the CTP formalism of non-local quantum field theory models reveal an instability.

\acknowledgments
It is pleasure to thank J\'anos Hajdu for several discussions.

\end{document}